\documentclass[superscriptaddress,onecolumn,secnumarabic,
amssymb,amsmath,nobibnotes,aps,prd,showkeys,showpacs,nofootinbib]{revtex4}

\usepackage[latin1]{inputenc}
\usepackage{graphicx}
\usepackage[english]{babel}

\usepackage{amsmath}
\usepackage{amssymb}
\usepackage{amsfonts}
\usepackage{colordvi}
\usepackage{psfrag}
\usepackage{color}

\begin{document}
\def\cL{{\cal L}}
\def\be{\begin{equation}}
\def\ee{\end{equation}}
\def\bea{\begin{eqnarray}}
\def\eea{\end{eqnarray}}
\def\beq{\begin{eqnarray}}
\def\eeq{\end{eqnarray}}
\def\tr{{\rm tr}\, }
\def\nn{\nonumber \\}
\def\e{{\rm e}}


\def\bef{\begin{figure}}
\def\eef{\end{figure}}
\newcommand{\ans}{ansatz }
\newcommand{\eeqn}{\end{eqnarray}}
\newcommand{\bd}{\begin{displaymath}}
\newcommand{\ed}{\end{displaymath}}
\newcommand{\mat}[4]{\left(\begin{array}{cc}{#1}&{#2}\\{#3}&{#4}
\end{array}\right)}
\newcommand{\matr}[9]{\left(\begin{array}{ccc}{#1}&{#2}&{#3}\\
{#4}&{#5}&{#6}\\{#7}&{#8}&{#9}\end{array}\right)}
\newcommand{\matrr}[6]{\left(\begin{array}{cc}{#1}&{#2}\\
{#3}&{#4}\\{#5}&{#6}\end{array}\right)}
\newcommand{\cvb}[3]{#1^{#2}_{#3}}
\def\lsim{\raise0.3ex\hbox{$\;<$\kern-0.75em\raise-1.1ex
e\hbox{$\sim\;$}}}
\def\gsim{\raise0.3ex\hbox{$\;>$\kern-0.75em\raise-1.1ex
\hbox{$\sim\;$}}}
\def\abs#1{\left| #1\right|}
\def\simlt{\mathrel{\lower2.5pt\vbox{\lineskip=0pt\baselineskip=0pt
           \hbox{$<$}\hbox{$\sim$}}}}
\def\simgt{\mathrel{\lower2.5pt\vbox{\lineskip=0pt\baselineskip=0pt
           \hbox{$>$}\hbox{$\sim$}}}}
\def\unity{{\hbox{1\kern-.8mm l}}}
\newcommand{\eps}{\varepsilon}
\def\ep{\epsilon}
\def\ga{\gamma}
\def\Ga{\Gamma}
\def\om{\omega}
\def\omp{{\omega^\prime}}
\def\Om{\Omega}
\def\la{\lambda}
\def\La{\Lambda}
\def\al{\alpha}
\newcommand{\ov}{\overline}
\renewcommand{\to}{\rightarrow}
\renewcommand{\vec}[1]{\mathbf{#1}}
\newcommand{\vect}[1]{\mbox{\boldmath$#1$}}
\def\tm{{\widetilde{m}}}
\def\mcirc{{\stackrel{o}{m}}}
\newcommand{\Dm}{\Delta m}
\newcommand{\dm}{\varepsilon}
\newcommand{\tanb}{\tan\beta}
\newcommand{\nbar}{\tilde{n}}
\newcommand\PM[1]{\begin{pmatrix}#1\end{pmatrix}}
\newcommand{\up}{\uparrow}
\newcommand{\down}{\downarrow}
\def\omE{\omega_{\rm Ter}}
%

\newcommand{\Dsusy}{{susy \hspace{-9.4pt} \slash}\;}
\newcommand{\DCP}{{CP \hspace{-7.4pt} \slash}\;}
\newcommand{\mc}{\mathcal}
\newcommand{\gr}{\mathbf}
\renewcommand{\to}{\rightarrow}
\newcommand{\gtc}{\mathfrak}
\newcommand{\wh}{\widehat}
\newcommand{\br}{\langle}
\newcommand{\kt}{\rangle}


\def\lsim{\mathrel{\mathop  {\hbox{\lower0.5ex\hbox{$\sim$}
\kern-0.8em\lower-0.7ex\hbox{$<$}}}}}
\def\gsim{\mathrel{\mathop  {\hbox{\lower0.5ex\hbox{$\sim$}
\kern-0.8em\lower-0.7ex\hbox{$>$}}}}}

\def\nn{\\  \nonumber}
\def\de{\partial}
\def\brf{{\mathbf f}}
\def\bbf{\bar{\bf f}}
\def\bF{{\bf F}}
\def\bbF{\bar{\bf F}}
\def\bA{{\mathbf A}}
\def\bB{{\mathbf B}}
\def\bG{{\mathbf G}}
\def\bI{{\mathbf I}}
\def\bM{{\mathbf M}}
\def\bY{{\mathbf Y}}
\def\bX{{\mathbf X}}
\def\bS{{\mathbf S}}
\def\bb{{\mathbf b}}
\def\bh{{\mathbf h}}
\def\bg{{\mathbf g}}
\def\bla{{\mathbf \la}}
\def\bmu{\mathbf m }
\def\by{{\mathbf y}}
\def\bmu{\mbox{\boldmath $\mu$} }
\def\bsig{\mbox{\boldmath $\sigma$} }
\def\bunity{{\mathbf 1}}
\def\cA{{\cal A}}
\def\cB{{\cal B}}
\def\cC{{\cal C}}
\def\cD{{\cal D}}
\def\cF{{\cal F}}
\def\cG{{\cal G}}
\def\cH{{\cal H}}
\def\cI{{\cal I}}
\def\cL{{\cal L}}
\def\cN{{\cal N}}
\def\cM{{\cal M}}
\def\cO{{\cal O}}
\def\cR{{\cal R}}
\def\cS{{\cal S}}
\def\cT{{\cal T}}
\def\eV{{\rm eV}}

\title{Quantum chaos inside Black Holes}

\author{Andrea Addazi}

\affiliation{ Dipartimento di Fisica,
 Universit\`a di L'Aquila, 67010 Coppito AQ, Italy}
 
 \affiliation{Laboratori Nazionali del Gran Sasso (INFN), 67010 Assergi AQ, Italy}

\date{\today}

\begin{abstract}
We show how semiclassical black holes can be reinterpreted as an effective geometry,
composed of a large ensamble of horizonless naked singularities (eventually smoothed at the Planck scale).
We call these new items {\it frizzyballs}, which can be rigorously defined by 
euclidean path integral approach. 
This leads to  interesting implications about information paradoxes. 
We demonstrate that infalling information will chaotically propagate
inside this system before going to the full quantum gravity regime (Planck scale). 


\end{abstract}

\maketitle

\section{Introduction}

Black holes' information paradox
is an insidious field of discussions. 
 Bekenstein-Hawking semiclassical approach \cite{Bekenstein,Hawking} suggests that,
from a pure state falling into a Black Hole, 
a highly mixed final state is obtained. This apparently implies that
 information is lost and unitarity is violated. On the other hand AdS/CFT correspondence 
seems to suggest that information is preserved in BH 
\cite{Maldacena:1997re,Witten:1998qj,Gubser:1998bc}
\footnote{An intriguing issue regarding holography is the locality. 
Recent discussion on effective non-local field theories were discussed in 
\cite{Addazi:2015dxa,Addazi:2015ppa}}. Complementarity in Black Hole physics \cite{Susskind:1993if,'tHooft:1984re}
is not enough to solve these problems: 
it leads to the firewall paradox \cite{Braunstein:2009my,Almheiri:2012rt}
 \footnote{Let us note that for extended theories of gravity, a problem could also be 
 the presence of instabilities just in the classical formulation of the theory.  
See \cite{Addazi:2014mga} for a discussion of external geodetic instabilities around BH or stars, in a class of massive gravity models. }. So, surely there is something more that we are missing 
in the complementarity picture.  
Information problem is "a battle field" of 
many different ideas and interpretations of Black holes
\cite{Dvali:2012gb,Dvali:2012rt,Dvali:2012en,Dvali:2014ila,Dvali:2015ywa,Foit:2015wqa,Casadio:2015bna,Casadio:2015sda,Hooft:2014daa,Carr:2015nqa,Lake:2015pma,Barrau:2014yka,Barrau:2015uca,Rovelli:2014cta,Saueressig:2015xua,DeLorenzo:2014pta,Giddings:2014nla,Giddings:2012gc,Germani:2015tda,Marolf:2015dia,Chatwin-Davies:2015hna,Susskind:2015toa,Afshordi:2015foa,Grudka:2015wva,Firouzjaee:2015bqa,Hawking:2014tga,Kawai:2014afa,Kawai:2013mda}
\footnote{The approach considered in \cite{Dvali:2012gb,Dvali:2012rt,Dvali:2012en,Dvali:2014ila,Dvali:2015ywa}
can be related to another class of classicalons studied in
\cite{Addazi:2014ila,Addazi:2015ata,Addazi:2015rwa,Addazi:2015hka,Addazi:2015eca,Addazi:2015fua,Addazi:2015oba,Addazi:2015goa,
Addazi:2015yna,Addazi:2015ewa}.}.

On the other hand, 
we know that in laboratory, entangled 
pairs of two photons can be prepared from ingoing pure photons.
This is the 
 so called
 {\it spontaneous parametric down-conversion} (SPDC) (see \cite{Horodecki:2009zz} for a review in these subjects):
 it consists of a nonlinear crystal pumped with 
 intense coherent light. 
 What happen is a three-wave mixing mechanism using  
(the lowest order) nonlinear susceptibility of non-linear birifrangent crystals. 
As well understood, this is not a violation of quantum mechanic principles:
the evolution from a pure state to a mixed one is just due
to a quantum decoherence effect, {\it i.e} 
the infalling information is entangled with the rest of the crystal
through interactions with it. 
In this case a wave function or an S-matrix approach 
describing the dynamics of the infalling information 
is not useful: a density matrix approach is preferred.
The density matrix associated to the infalling pure state has a non-unitary 
evolution described by a Liouville equation. 
Following such an analogy, we would like to suggest 
that a mechanism similar to SPDC converts infalling pure states into out-falling mixed ones  inside 
quantum black hole, without violating unitarity at fundamental level.
In this case, a quantum decoherence mechanism is realized. 
On the other hand, such a mechanism seems not 
possible in a semiclassical black hole: its geometry is smoothed 
and a particle infalling into a semiclassical BH will not experience 
drastic changes because of the equivalence principle. 


In this paper, we suggest that 
a semiclassical black hole geometry is an approximated 
effective solution: it is 
  a system of a large number of horizonless geometries, in semiclassical regime.
Such a system can be imagined as a sort of "frizzy-ball":
space-time asperities will differentiate the 
ideal semiclassical black hole by the complicated 
system of metrics. A "frizzy black hole"
has an approximated classical event horizon,
approximately emitting Bekenstein-Hawking radiation. 
How much "frizzy" with respect to a Semiclassical Black hole 
can be rigorously defined by the departure of 
its emitted radiation with respect to a Bekenstein-Hawking thermal
one, as we will discuss later. However, also a frizzyball
emits a mixed state rather than a pure one for $t<<t_{BH}$. 
\footnote{However, the effect of environmental radiation infalling into BH,
such as CMB, was studied in \cite{Firouzjaee:2015bqa}. They seem to demonstrate that
environmental radiation can relevantly affect and suppress the emission 
of Bekenstein-Hawking radiation. Also in frizzy-balls, this effect is expected 
to be similar. }. 


This suggestion has important implications in 
the evolution of infalling information.  
Let us consider an initial pure state infalling into
a frizzy black hole. What one could expect is that 
initially this just "feels" an approximately smoothed semiclassical space-time.
In fact the quantum wave function of an infalling particle  has a De Broglie 
wave lenght
that
is much larger than asperities' scales. 
However, inevitably, the infalling wave will start to be blueshifted,
so that it will start to "resolve" more and more the asperities of the non-trivial topology. 
At this point, the infalling wave will start to be scattered back and forth
by the asperities, before going to the Planck scale. 
At this stage, information 
will start to be chaotized inside the system \footnote{
For different applications of chaos theory in information paradox
see \cite{Hawking:2014tga,C0,Polchinski:2015cea,C1}.}.
This system can be thought as a wave function scattering on a {\it quantum Sinai billiard}!
 \footnote{Quantum chaos studies the relationship
 between classical chaos and quantum mechanics.
 Quantum Sinai Billiard is a well known example 
 of quantum chaotic system. 
 For a general overview in quantum chaos see \cite{QuantumChaos}.}

 What one will expect is that the initial probability will be fractioned into
two contributions. In fact, a part of the initial probability density will "escape", emitted as quasi-thermal 
radiation, 
by the system while a part will remain "trapped" forever in the system 
because of back and fourth scatterings. 
This can be easily understood by a classical chaotic mechanics point of view.
In fact, the definition of a classical chaotic scatterings  of a particle is the following:
a classical mechanics' scattering problem in which 
the incident particle can be trapped ideally forever 
in a class of classical orbits; but the periodic orbits are
unstable saddle solutions and their number grows exponentially 
with time. 
Chaotic scatterings have a high sensitivity to the initial conditions
manifesting itself in a fractal chaotic invariant set, which is also 
called {\it chaotic saddle} \cite{QCS}.
Energy shells closed to the chaotic saddle energy shell
will continue to be chaotic. 
\footnote{An example in simpler systens are Kolmogorov-Arnold-Moser (KAM) elliptic islands, 
that contain stable periodic orbits.
KAM stable periodic orbits undergo to chaotic bifurcations, 
rupturing the smoothed topology of the invariant 
set \cite{QCS2}. As usually happening for chaotic saddles, 
KAM islands are surrounded by a layer of chaotic trajectories. 
Another typical example is the hyperbolic set of hyperbolic unstable 
trajectories: solutions are exponentially growing or decreasing but the
number of directions are constants of motion. }.
In our case,  periodic orbits will be 
forever trapped in back and forth scatterings
among the the space-temporal Sinai biliard. As generically happening in 
classical chaotic scatterings' problems,
these trajectories will necessary exist in the phase space of the system
\footnote{
Our problem is nothing but a complication with respect to
a simpler and well known example of classical chaotic scattering 
problem: a 2d classical elastic scattering of
a particle on a system of N fixed disks of radius $a$
\cite{QCS3}.
In this simple problem, kinetic energy is assumed to be conserved, 
{\it i.e} no any dissipations are considered. 
For one disk the problem is trivially un-chaotic:
the differential cross section is just $\frac{d\sigma}{d\theta}=\frac{a}{2}|\sin \frac{\theta}{2}|$
for $\theta$ in the range $[-\pi,\pi]$;
and no trapped periodic orbit are possible. 
However, with two disks, an unstable periodic orbit is the one 
bouncing back and forth forever among the two disks.
With the increasing of the number of disks one can easily 
get that the number of trapped periodic orbit will exponentially 
increase. 
For example, as shown in \cite{QCS4}, 
in a three disks' system, the 
number of unstable periodic orbits proliferate as $2^{n}$
where $n$ is the number of bounces in unit of the period.
if the radius is the distance among the 
next neighboring disk is $R>2.04822142\, a$. }.
From Classical chaotic scatterings' 
one can get the main feature of the quantum semiclassical 
chaotic problem associated and about semiclassical 
periodic orbits. 
So, 
because of multiple diffractions and back and fourth scatterings, 
one will also expect that the resultant wave function is "chaotized" by the system:
the total wave function is a superposition of the initial one 
plus all the spherical ones coming from each "scatterators".  
A part of the initial infalling information will be trapped "forever" 
in the system, {\it i.e} for all the system life-time.
In order to describe the evolution of the infalling informations,
a quantum mechanical approach 
based on wave functions  is not useful, in this system.
A wave functions' approach can be substituted by a quantum statistical mechanics' 
approach in terms of density matrices. 
From the point of view of a Quantum field theory, a S-matrix approach is not useful in this case,
even if "fundamentally true":
in order to calculate $\langle in|S| out \rangle$ ($in$ is the in-going plane wave,
where $out$ is the out-going result), one has to 
get unknown informations on the precise geometric configuration 
inside the system and about the trapped information state inside it. 
Such a system can emit a quasi thermalized mixed information state
without losing any informations at fundamental level. 
In other words, we 
 suggest that the space-time non-trivial topology prepares an entangled 
state as well as an experimental apparatus can prepare 
an entangled state by an initial pure state. 
In our case, the effect will also be dramatically efficient: thank to quantum field theory' interactions in the lagrangian density functional, n-wave mixings will occur
inside our system. Thinking about the ingoing state as a collection 
of coherent quantum fields, these will be scattered into 
the system and, they will meet each others inside "the trap", they will scatter each others,
coupled by lagrangian interactions.
A complicated cascade of hadronic and electromagnetic processes
is expected. For example, these will produce a large amount of neutral pions, 
that will electromagnetically decay into two entangled photons 
$\pi^{0}\rightarrow \gamma \gamma$ ($\tau\simeq 10^{-16}\, \rm s$ in the rest frame). 
However, also from only one plane wave infalling in the system, 
the final state emitted by the system will be a mixed state:
this is just an effect of the information losing inside the system
because of trapped chaotic zones inside.
This phenomena is a new form of quantum decoherence  
induced by the space-time topology. Usually, quantum decoherence is
the effective losing of infalling informations in a complex system,
like coherent light pumped in a non-linear crystal. 
In this case,  the complex topology of space-time  
catalyzes the effective losing of information. 

In our quantum chaotic system, we will not have any information paradoxes or firewalls.
In fact, infalling pure information is converted to a mixed thermal state
during $t<<t_{BH}$ because of a quantum decoherence  
inside the space-time Sinai biliard. The evolution of this state is apparently 
non-unitary, but unitarity is fundamentally preserved: 
the lost information is trapped in chaotic zones inside the biliard.
However, it seems that such a system cannot really hide information 
"forever": it has approximately the same Hawking's radiation (same thermal entropy) of BH
for $t<<t_{BH}$,
so that it will completely evaporate after a certain time. 
As a consequence, during the final evaporation, 
 the hidden information cannot be trapped anymore and 
 it is re-given to the external environment 
 in a "final information burst".
So, CPT is "apparently" violated for a time $t< t_{BH}$
in the external environment,  but,
after the final evaporation, CPT again manifests 
its conservation \footnote{Another apparent violation of CPT induced 
in neutron-antineutron system by a new interaction  was discussed in \cite{Addazi:2015pia}.}. 
Such a phenomena is a sort of space-time phase transition,
defined as a transition of the space-time topology 
itself: the frizzy space-time will transmute to a 
Minkowski like one.  
The trapped probability density is expected to be linearly dependent by
the number of asperities as $\rho \sim N_{s}e^{-\Gamma T}$,
where $\Gamma$ is proportional to effective average deepness of 
systems of asperities trapping $\rho$. (This can be easily estimated in WKB 
approach). 
On the other hand, the number of asperities are sustained by the Black hole mass, 
in turn decreasing with the temperature as $dM/dT=-1/8\pi T^{2}$. 
As a consequence, $\rho$ is approximately described by
a simple differential equation 
$d\rho(T)/dT\sim -\frac{1}{T^{2}}e^{-\Gamma T}$
\footnote{In order to avoid any confusions, let us clarify 
what we mean for (quasi) "forever" trapped information.
In fact, one observer outside and one inside the frizzyball 
will disagree about BH lifetime. The disagreement depends by 
the BH mass, but typically for $M>M_{\odot}$ the observer
inside the frizzyballs will measure $t_{BH}<<1\, \rm yr$. 
However, this short life-time is enough for an efficient 
chaotization: particles' waves get high kinetic energy by the gravitational field, {\it i.e} we are considering a relativistic Sinai Biliard, with very fast quantum interactions 
among fields. }

The paper is organized as follows:
In Section 2, we argument how a system of N horizonless singularities  can recover 
a Semiclassical BH state; in Section 3 we will discuss the problem 
the problem 
of chaotic scatterings of matter infalling in a frizzyball,
in Section 4 we show our conclusions. 

\section{The path-integral approach}

In this section, we will give 
a path-integral formulation of our problem. 
We will precisely define what is a horizonless "frizzy-ball"
with respect to a semiclassical euclidean black hole. 

In general,
the path integral over all euclidean metrics and matter fields
is 
\be \label{formaly}
Z_{E}=\int \mathcal{D}g\mathcal{D}\phi e^{-I[g,\phi]}
\ee
where $g$ is the euclidean metric tensor. 
In semiclassical approach the relevant generally covariant lagrangian \cite{Euclidean}
is 
\be \label{I}
I_{E}=-\int_{\mathcal{M}}\sqrt{g}d^{4}x\left(\mathcal{L}_{m}+\frac{1}{16\pi}R\right)+\frac{1}{8\pi}\int_{\partial \mathcal{M}}\sqrt{h}d^{3}x(K-K^{0})
\ee
(we use $G_{N}=1$),
where $K$ is the trace of the curvature induced on the boundary $\partial \mathcal{M}$
of the region $\mathcal{M}$ considered, $h$ is the metric induced on the boundary $\partial \mathcal{M}$,
$K^{0}$ is the trace of the curvature induced imbedded in flat space.
The last term is a contribution from the boundary.
As usually done in semiclassical WKB approach, 
one can perturb matter fields and metric
as $\phi=\phi_{0}+\tilde{\phi}$
and $g=g_{0}+\tilde{g}$, 
so that 
$$I[\phi,g]=I[\phi_{0},g_{0}]+I_{2}[\tilde{\phi},\tilde{g}]+higher\,orders$$
$$I_{2}[\tilde{\phi},\tilde{g}]=I_{2}[\tilde{\phi}]+I_{2}[\tilde{g}]$$
\be \label{Z}
log Z=-I[\phi_{0},g_{0}]+log \int \mathcal{D}\tilde{\phi} \mathcal{D}\tilde{g} e^{-I_{2}[\tilde{g},\tilde{\phi}]}
\ee

In an Euclidean Schwarzschild solution, 
the metric has a periodicity in time $i\beta$,
where 
$$\beta=T^{-1}=8\pi M$$
with $T,M$ BH temperature and mass. 
This metric has a form 
\be \label{form1}
ds_{E}^{2}=\left(1-\frac{2M}{r}\right)d^{2}\tau+\left(1-\frac{2GM}{r}\right)dr^{2}+r^{2}d\Omega^{2}
\ee
but this can be conveniently rewritten 
in terms of a new variable
$$x=4M\sqrt{1-\frac{2M}{r}}$$
\be \label{ES}
ds_{E}^{2}=\left(\frac{x}{4M}\right)^{2}+\left(\frac{r^{2}}{4M^{2}}\right)^{2}dx^{2}+r^{2}d\Omega^{2}
\ee
that it is free by the mathematical singularity in $r=2M$,
while the Euclidean time $\theta$ is angular variable with period
$\beta=8\pi M$. As a consequence, the boundary 
$\partial M$ has a topology $S^{1}\times S^{2}$
at a certain fixed radius $r_{0}$. 
In a stationary-phase point approximation, 
the path integral becomes just 
a partition function of a canonical ensamble, 
with temperature $T=\beta^{-1}$. 
In semiclassical approximation, 
the dominant contribution to the path integral in a Euclidean background is 
\be \label{dom}
Z_{ES}= e^{-\frac{\beta^{2}}{16\pi}}
\ee
Such a term is coming by surface integrals in the action. 

(\ref{dom}) is thermo-dynamically related to the average energy as
\be \label{average}
\langle E \rangle=-\frac{d}{d\beta}(log Z)=\frac{\beta}{8\pi} 
\ee
while $log Z$ is usually defined in statistical mechanic
as 
\be \label{usuallyd}
\rm log Z=-\frac{F}{T}
\ee
But 
the entropy is related to the average and free energy as
\be \label{SFU}
S=\beta(F-\langle E \rangle)
\ee
so that one can obtain the Hawking's entropy 
\be \label{S1}
S=\beta(log Z-\frac{d}{d\beta}(log Z))=\frac{\beta^{2}}{16\pi}=\frac{1}{4}A
\ee

Let us remember that the physical interpretation 
of the semiclassical BH path integral is
that a BH is confined in a box 
with a fixed size, 
and it is consider in thermal equilibrium with 
its own Hawking radiation, at a constant temperature T.

After this short review, let us give the definition of frizzyball.

{\bf Def}:
let us consider a generic system of  N horizonless singularities (suppose to be eliminated at the Planck scale)
inside a box with a surface $\partial \mathcal{M}$.
 This system is a {\it frizzyball} if it satisfies the following hypothesis:
 
 i) The N horizonless singularities are  
 in thermal equilibrium with the box, and 
 a formal definition of partition functions $Z_{I}$ for each metric tensor $g^{I=1,...,N}$
 exists.
 
 ii) In semiclassical approximation, the leading order of the total partition 
function associated to this system is the product of the single partition function:
\be \label{Impl1}
Z_{TOT}= \prod_{I=1}^{N} Z_{I}
\ee
This corresponds to consider the total entropy in the system as the sum of entropies associated 
to each naked singular geometries, {\it i.e}
\be \label{Impl2}
log Z_{TOT}= \sum_{I=1}^{N}log Z_{I}
\ee
The physical interpretation is that the 
intergeometries' interactions are negligible 
with respect to the temperature of the system inside the box
\footnote{This approximation seems not compatible 
with chaotization of information inside the frizzyball. 
Infact, chaotization is related to an exchange of matter 
among the geometries. However, this apparent contradiction is avoid in frizzyball system.
In fact, if the net exchange of 
heat among the geometries is negligible with respect to the 
thermal energy in the box, as expected 
in a system in thermal equilibrium, this approximation will be rightly applied. 
It corresponds to $S_{int}\simeq 0$. As regards gravitational interactions among the metrics, 
this is strongly suppressed in semiclassical regime. }.

iii) The total average partition function is
\be \label{approx}
\langle Z_{TOT}\rangle=e^{-\frac{\beta^{2}}{16\pi}-\frac{\sigma_{\beta}^{2}}{16\pi}}=Z_{ES}e^{\frac{-\sigma_{\beta}^{2}}{16\pi}}
\ee
where $\sigma_{\beta}$ is the variance of $\beta$-variable in the system,
and it is assumed to be very small even different from zero.
In fact this parametrize the small deviations of the semiclassical 
frizzy-ball with respect to semiclassical BH, {\it. i.e} the local not perfect smoothness 
of the frizzy geometry. 
Eq.(\ref{approx}) is understood considering deviations $\beta+\delta \beta$, with $\delta \beta<<\beta$:
$$e^{-\frac{\beta^{2}+2\beta \delta \beta+O(\delta \beta^{2})}{16\pi}}
=Z_{ES}\times e^{\frac{\beta \delta \beta}{8\pi}+O(\delta \beta^{2})}$$
and assuming $\langle \delta \beta \rangle=0$. 

Eq.\ref{approx} leads to the entropy 
\be \label{finalentorpy}
\langle S \rangle=\frac{\beta^{2}}{16\pi}+\frac{\sigma_{\beta}^{2}}{16\pi}
\ee
Let us note that even if a small correction to the Bekenstain-Hawking entropy is predicted,
the out-going radiation is expected to be a mixed state also for this system. 

The next non-trivial step is to demonstrate the mathematical consistence 
of the definition of a frizzy-ball, {\it i.e} if the three hypothesis not lead to 
any contractions. 
In semiclassical approximation, 
the existence of a frizzy-ball is related to the following identity:
\be \label{Impl2}
-I[g_{0},\phi_{0}]+log \int \mathcal{D}\tilde{\phi}e^{-I_{2}[g_{0},\mathcal{\phi}]}+log \int\mathcal{D}\tilde{g}e^{-I[\tilde{g}]}= -\sum_{J}I[g_{0}^{J},\phi_{0}]+	\sum_{J}\left[log \int \mathcal{D}\tilde{\phi}e^{-I_{2}[g_{0}^{J}\mathcal{\phi}]}+log \int\mathcal{D}\tilde{g}^{J}e^{-I[\tilde{g}^{J}]}\right]
\ee
where $g_{0}$ is the Euclidean Schwarzschild metric tensor
while $g_{0}^{J}$ are the Euclidean metric tensor of $J=1,...,N$ geometries. 
This leads to the following classical relations 
\be \label{right1}
g_{0}= (\sum_{J}\sqrt{g_{0}^{J}})^{2}+higher\,orders
\ee
\be \label{right2}
\sqrt{g_{0}}R(g_{0}) =\sum_{J}\sqrt{g_{0}^{J}}R(g_{0}^{J})+higher\, orders
\ee
while quantum fluctuations are 
\be \label{right3}
I_{2}[\tilde{g}]=\sum_{J}I_{2}[\tilde{g}^{J}]+higher\,orders
\ee

\section{"Black" Chaotic Sinai Biliard}
In this section, we will argument
the apparently information lost 
in a system of horizonless singularities.
In subsection 3.1 and 3.2 we will use a non-relativistic approach.
This approximation is not fully justified in our realistic problem, 
as well as a non-relativistic quantum mechanic approach 
to scattering problems in particle physics. 
However, one could retain useful to discuss simplified problems
rather than the realistic one,
in order 
to get easier relevant chaotic aspects. 
In subsection 3.3 we will formally comment our problem from 
a QFT point of view. 

\subsection{Classical chaotic scattering on a Space-time Sinai Biliard}

The classical chaotic scattering of a particle on 
an box of horizonless singularities 
is characterized by a classical Hamiltonian system 
$\dot{\vect{r}}=\partial H/\partial \vect{p}$ 
and $\dot{\vect{p}}=-\partial H/\partial \vect{r}$
with an initial condition $x_{0}=(\vect{r}_{0},\vect{p}_{0})$
in the space of phase. 
In particular, one can define N Hamiltonian 
systems for each geometry, describing the
motion of the particle on each of N geometries.
Clearly, one can obtain similar systems 
by the geodesic equations 
$\ddot{x}^{\mu}+\Gamma_{\alpha\beta}^{\mu} \dot{x}^{\alpha}\dot{x}^{\beta}=0$
of the particles in each $g^{I}_{\mu\nu}$ metrics,
{\it i.e} the propagation of the particle 
on the Ith hypersurfaces. 
The effective 
Hamiltonian obtained for the propagation 
on a Ith metric is 
$$H_{I}=\frac{1}{2m}p_{i}g^{ij}p_{j}=\frac{p^{2}}{2m}+V$$
in non-relativistic regime, 
where $V$ is an effective potential depending on the Ith geometry. 
The solution of such a system will be determined by 
a trajectory 
$x_{t}=\phi^{t}(x_{0})$ solving the Chauchy problem of classical mechanics.
In this case, we will expect a 
proliferation of trapped periodic unstable trajectories,
as anticipated in the introduction, 
because of an infinite back and forth 
scattering among the N geometries. 

Let us define the action of the classical problem:
\be\label{action}
S(E)=-\int_{\Sigma} \vect{r}\cdot \vect{p}
\ee
where $\Sigma$ is the energy shell $H=E$
where a scattering orbit is sited. 
The time delay is defined 
as 
\be \label{time}
\mathcal{T}(E)=\frac{\partial S}{\partial E}
\ee
If the impact parameters of the initial orbits $\rho$ 
has a probability density $w(\rho)$, 
the probability density conditioned by energy $E$ of the corresponding 
time delays is 
\be \label{Prob}
P(\tau|E)=\int d\rho w(\rho)\delta(\tau+\mathcal{T}(\rho|E))
\ee
where the condition "corresponding time delays" is encoded
in the integral though the Dirac's delta. 
(\ref{Prob}) is useful to describe the escape of the particle from the 
trapped orbits' zone. 
Inspired by $N$ disks' problems studied in literature, 
an hyperbolic invariant set is expected to occur.
In this case the decays' distribution rate is 
expected to exponentially decrease, {\it i.e}
\be \label{expdcr}
lim_{t\rightarrow \infty}\frac{P(\tau|E)}{t}=-\gamma(E)
\ee
On the other hand, for non-hyperbolic sets, 
like KAM elliptic islands, 
power low decays
are generically expected 
$P(t|E)\sim 1/t^{\alpha}$,
where $\alpha$ depends by 
the articular density of trapped orbits.

Now, let us discuss the time delays in our system of horizonless geometries. 
If unstable periodic orbit exists in our scattering problem,
eq.(\ref{time}) will have $\rho$-poles,
{\it i.e} it becomes infinite for  
precise initial impact parameters $\rho$. 
Let starts with the simplest case
of a scattering on one geometry.
Let us suppose that this geometry has 
$n$-periodic directions. 
For example, a conic singularity has a periodic direction 
around its axis and so on. 
In this case, 
the 
integral (\ref{time}) 
has a couple of asymptotic divergent 
direction along each paths 
$x_{t}^{\theta^{(n)}}=(\theta^{(n)},p_{\theta}^{(n)})$,
where $\theta^{(n)}$ are the periodic variables and 
$p_{\theta}^{(n)}$ are their conjugated momenta. 
The particle will be infinitely trapped in these
paths if and only if its 
initial incident direction is 
parallel to one of the periodic directions $\theta^{(n)}$.

Now let us complicate the problem considering two geometries.
In these case the number of divergent asymptotes of $\mathcal{T}$
correspond to three couples: 
i) cycles around the first geometry, ii) cycles around the second geometry,
iii) trapped back and forth trajectories between the two geometries. 
As a consequence, just in this case the number of trapped trajectories
is enormously growing. 

One can easily get that for a N number of horizonless singularities 
the number of the divergent asymptotes 
for the time-delay function
will proliferate. 
These divergent asymptotes are connected to the fractal character 
of the invariant set. 
A geometric way to see the problem is the following:
one can consider a $2\nu-2$ Poincar\'e surface with section in the Hamiltonian
flown on a fixed energy surface, where $\nu$ is the number of degree of freedom
of the system. In our case, we consider a 4d Poincar\'e surface. 
The time-delay of the orbit necessary to go-out 
from the cones at large enough distances 
is $\mathcal{T}_{\pm}(\rho|E)$, for every initial 
Chauchy condition in the Poincar\'e section. 
 $\mathcal{T}_{+}(\rho|E)\rightarrow \infty$ 
 stable surfaces of orbits trapped forever.
 On the other hand, $\mathcal{T}_{-}\rightarrow \infty$
 on unstable manifolds of orbits 

 In other words, $|\mathcal{T}_{-}(\rho|E)|+|\mathcal{T}_{+}(\rho|E)|$
 is a localizator functions for the fractal set trapped trajectories.
 
Let us remind the definition of sensitivity to initial conditions,
defined by the Lyapunov exponents 
\be \label{Lyapunov}
\lambda(x_{0}|\delta x_{0})=lim_{t\rightarrow \infty}\frac{1}{t}\frac{|\delta x_{t}|}{|\delta x_{0}|}
\ee
where $\delta x_{0,t}$ are infinitesimal perturbation of the 
initial condition $x_{0}$ and the resultant 
orbit $x_{t}$. 
In general, the Lyapunov exponents depend on the initial perturbation 
and on the orbit perturbation.
However, $\lambda$ becomes un-sensible by the orbit
in ergodic invariant sets 

These sets are characterized by 
the following hierarchy of Lyapunov exponents
in a system with $\nu$-degrees of freedom:
\be
\label{lambdaf0}
0=\lambda_{\nu}\leq \lambda_{\nu-1} \leq .... \leq \lambda_{2} \leq \lambda_{1}
\ee
while
\be
\label{lambdaf1}
0=\lambda_{\nu+1}\ge ... \ge \lambda_{2\nu}
\ee
In a Hamiltonian system, 
the symplectic flows of the Hamiltonian 
operator implies 
that 
$$\sum_{k=0}^{2\nu}\lambda_{k}=0$$
and 
$$\lambda_{2\nu-k+1}=-\lambda_{k}$$
where $k=1,2,...,2\nu$. 
In our case, the number of degree of freedom 
is $\nu=3$, so that the number of
independent Lyapunov's exponents characterizing 
the chaotic scattering is three. 

The exponentially growing number of unstable
periodic trajectories inside the invariant set 
is characterized by a topological number 
\be \label{topn}
h=lim_{t\rightarrow \infty}\frac{1}{t}ln(\mathcal{N}\{\tau_{o}\geq t\})
\ee
where $\mathcal{N}$ is the number of 
periodic orbits of period minor than $t$, $\tau_{o}$ is the periodic orbit time. 
Such a number is the so called topological entropy

$h>0$ if the system is chaotic
while $h=0$ if non-chaotic. 
For a system like 
a large box of horizonless geometries, 
this number will be infinite. 
Such a number will diverge just with only three geometries
as happen just in a system of three 2d-disks. 

In our system, as for disks, a  
 hyperbolic invariant set or something of similar is expected.  
For this set $\delta V$ small volumes
are exponentially 
stretched 
by  
\be \label{factorgrowing}
g_{\omega}=exp\{\sum_{\lambda_{k}>0}\lambda_{k}t_{\omega}\}>1
\ee
because of its unstable orbits; 
where $t_{\omega}$ is the time interval associated
to the periodic orbit of period $n$,
{\it i.e} to the symbolic 
dynamic $\omega=\omega_{1}....\omega_{n}$
corresponding to 
all the nonperiodic 
and periodic orbits 
remaining closed 
in a $\delta V$ for a time
$t_{\omega}$.
Using (\ref{factorgrowing}),
one can 
weight the probabilities for
trapped orbits as 
\be \label{probweight}
\mu_{\alpha}(\omega)=\frac{|g_{\omega}|^{-\alpha}}{\sum_{\omega}|g_{\omega}|^{-\alpha}}
\ee
This definition is intuitively understood:
a highly unstable
trajectory with $g_{\omega}>>1$ is 
weighted as $\mu_{\alpha}\simeq 0$.
The definition (\ref{probweight}) is normalized 
$\sum_{\omega}\mu_{\alpha}(\omega)=1$.
With $\alpha=1$ we recover the ergodic definition
for the Hamiltonian 
system.

An intriguing question will be if one can determine the 
Hausdorff dimension of the fractal sets
for our box of cones. 
In principle, the answer is yes, but in practice 
the problem seems really hard to solve.
In order to get the problem let us define 
the Ruelle topological pressure 

 \be \label{Ruelle}
 P(\alpha)=lim_{t\rightarrow \infty}\frac{1}{t}ln \sum_{\omega, t<t_{\omega}<t+\Delta t}|g_{\omega}|^{-\alpha}
 \ee
Ruelle topological pressure is practically independent by $t,\Delta t$ for 
a large 
$\Delta t$. 
The Ruelle topological pressure has a series of useful relations:

1) $P(\alpha_{1}+\alpha_{2})\leq P(\alpha_{1})+P(\alpha_{2})$

2) $P(0)=h$, {\it i.e} for $\alpha=0$ the Ruelle topological pressure
is just equal to the topological entropy.

3) $P(1)=-\gamma$, {\it i.e} for $\alpha=1$ the Ruelle topological 
pressure is just equal to the escape rate.

4) The Ruelle topological pressure is connected to Lyapunov's exponents as
$$\frac{dP}{d\beta}(1)=-lim_{t\rightarrow \infty}\sum_{\omega,t<t_{\omega}<t+\Delta t}
\mu_{1}(\omega)ln|g_{\omega}|=-\sum_{\lambda_{k}>0}\lambda_{k}$$

The last relation is the one 
connecting the Ruelle topological pressure with the Hausdorff
dimension $d_{H}$: 
5) $P(d_{H})=0$.
The Hausdorff dimension of a system with $\nu$ d.o.f is 
bounded as $0\leq d_{H} \leq \nu-1$
for the subspace of unstable directions, 
while a corresponding set of stable directions 
has exactly the same dimension of the previous one. 
Let us note that for a system with $\nu=1$
the Hausdorff dimension will collapse
to $d_{H}=0$, {\it i.e} no chaotic dynamics. 
In our case, $0\leq d_{H} \leq 2$
and in principle it can be founded 
as a root of the Ruelle topological pressure. 

\subsection{Semiclassical chaotic scattering on a Space-time Sinai Biliard}

A natural approach to quantum chaotic scattering can be to 
consider a semiclassical approach correspondent to the classical chaotic problem.
In Semiclassical approach, 
the main aspects of fully classical limit 
are remaining: trapped periodic orbits, invariant sets
and so on. 
In semiclassical approach one can 
generalize the classical notion of time delay 
for a semiclassical quantum system. 

Let us remind, just to fix our conventions,  that $\psi_{t}(\vect{r})$ 
is obtained by an initial $\psi_{0}(\vect{r}_{0})$ 
by the unitary evolution 
\be \label{psitpsi}
\psi_{t}(\vect{r})=\int d\vect{r}_{0}K(\vect{r},\vect{r}_{0},t)\psi_{0}(\vect{r}_{0})
\ee
where $K$ is the propagator, represented as a non-relativistic Feynman path integral 
as 
\be \label{Krrz}
K(\vect{r},\vect{r}_{0},t)=\int \mathcal{D}\vect{r} e^{\frac{i}{\hbar}I}
\ee
where
$$I=\int_{0}^{t} dt L(\vect{r},\dot{\vect{r}})$$
I the action and $L$ the lagrangian of the particle. 
The semiclassical limit is obtained in the limit 
$$I=\int_{0}^{t}[\vect{p}\cdot d\vect{r}-Hd\tau]>>\hbar$$
so that the leading contribution to the path integral 
is just given by classical orbits. 
The corresponding WKB propagator
has a form 
\be \label{WKB}
K_{WKB}(\vect{r},\vect{r}_{0},t)\simeq \sum_{n}\mathcal{A}_{n}(\vect{r},\vect{r}_{0},t)e^{\frac{i}{\hbar}I_{n}}
\ee
where we are summing on all over the classical orbits of the system, 
while amplitudes $\mathcal{A}_{n}$
are 
\be \label{amplitudesss}
\mathcal{A}_{n}(\vect{r},\vect{r}_{0},t)=\frac{1}{(2\pi i \hbar)^{\nu/2}}\sqrt{|det[\partial_{\vect{r}_{0}}\partial_{\vect{r}_{0}}I_{n}[\vect{r},\vect{r}_{0},t]]|}e^{-\frac{i\pi h_{n}}{2}}
\ee
where $h_{n}$ counts the number of conjugate points along the n-th orbit. 

From (\ref{amplitudesss}), the probability amplitude is related to Lyapunov exponents 
as 
\be \label{unstableAA}
|\mathcal{A}_{n}|\sim exp\left(-\frac{1}{2}\sum_{\lambda_{k}>0}\lambda_{k}t \right)
\ee
along unstable orbits.
On the other hand, 
\be \label{stableSSS}
|\mathcal{A}_{n}|\sim |t|^{-\nu/2}
\ee
along stable orbits

The level density of bounded quantum states 
is described by the trace of the propagator.
In semiclassical limit, 
the trace over the propagator
is peaked on 
around the periodic orbits 
and stationary saddle points. 
This allows to semiclassically quantize 
semiclassical unstable periodic orbits
that are densely sited in the invariant set.
As a consequence, the 
semiclassical quantum time delay is 
\be\label{timed}
\mathcal{T}=\int \frac{d\Gamma_{ph}}{(2\pi\hbar)^{\nu-1}}[\delta(E-H_{0}+V)-\delta(E-H_{0})]+O(\hbar^{2-\nu})+2\sum_{p}\sum_{p} \tau_{a=1}^{\infty}\tau_{p}\frac{\cos \left(a\frac{S_{p}}{\hbar}-\frac{\pi a}{2}{\bf m}_{p}\right)}{\sqrt{|det(\mathcal{M}_{p}^{a})|}}+O(\hbar)
\ee
where $d\Gamma_{ph}=d\vect{p}d\vect{r}$
and  the sum is on all the periodic orbits (primary periodic orbits $p$ and the
number of their repetitions $a$);
$S_{p}(E)=\int \vect{p}\cdot d\vect{r}$, $\tau_{p}=\int_{E}S_{p}(E)$,
${\bf m}_{p}$ is an index called Maslov index, 
and $\mathcal{M}$ is a $(2\nu-2)\times (2\nu-2)$ matrix 
associated to the Poincar\'e map in the 
neighborhood of the a-orbit.
A geodetic equation of a particle on a geometry can be mapped to a problem with an Hamiltonian interaction $V$,
as done in (\ref{timed}).

Now, let us consider a simplified problem with only $\nu=2$
d.o.f, in order to more easily get analytical important proprieties 
of semiclassical chaotic scatterings and their features.
Let us consider a generic projection of our box of geometries
to a 2d plane. Now, we study the dynamics
in this plane, ignoring the existence of a third dimension.
However, we can be so general in our consideration 
to be practically valid for every chosen projection!
Clearly, we remark that we know well how this problem 
can be only a different simplified problem with respect 
the 3d one. 
In this case, the matrix $\mathcal{M}$ has two eigenvalues:
$\{g_{p},g_{p}^{-1} \}$, where $g_{p}$ is the classical factor
$|g_{p}|=exp(\lambda_{p}\tau_{p})$.
As a consequence the complicate equation (\ref{timed})
for the time delay is just reduced 
to 
\be \label{TEred}
\mathcal{T}(E)=\mathcal{T}_{0}(E)-2\hbar Im\frac{d\,ln\,Z(E)}{dE}+O(\hbar)
\ee
where $\mathcal{T}_{0}(E)$ is the analytical part 
given by the first integral in (\ref{timed}),
while $Z(E)$ is the Zeta function 
\be \label{Zfu}
Z(E)=\prod_{p} \prod_{a=0}^{\infty}\left(1-e^{ia\phi_{p}}\frac{1}{g_{p}^{a}\sqrt{|g_{p}|}}\right)
\ee
where 
$$\phi_{p}=\frac{1}{\hbar}S_{p}-\frac{\pi}{2}{\bf m}_{p}$$
From (\ref{TEred}) and (\ref{Zfu}) one could get,
as an application of the Mittag-Leffler theorem, 
that 
the pole of the resolvent operators 
exactly corresponds to the zeros of the 
Zeta function. In complex energies' plane, 
the contribution of periodic orbits to 
the trace of the resolvent operator 
is related to the Z function by the simple relation
\be \label{relasimple}
tr \frac{1}{z-H}|_{p}= \frac{d}{dz}lnZ(z)=\frac{1}{i\hbar}\sum_{p}\sum_{a}\tau_{a}e^{ia\phi_{p}}\frac{1}{|g_{p}|^{a/2}}
\ee
(we omit extra higher inverse powers of $|g_{p}|$).
But the poles of the resolvent operator and the zeros of the Zeta function 
are nothing but scattering resonances:
$$Z(E_{a}=\mathcal{E}_{a}-i\Gamma_{a}/2)=0$$
Let us comment that if the invariant set contains
a single orbit, 
resonances $E_{a}$
satisfy the Bohr-Sommerfeld quantization condition
$$S_{p}(\mathcal{E}_{a})=2\pi \hbar \left( a+\frac{1}{4}{\bf m}_{p}\right) +O(\hbar^{2})$$
while widths satisfy
$$\Gamma_{a}=\frac{\hbar}{\tau_{p}}\,ln\,|g_{p}(\mathcal{E}_{r})|+O(\hbar)$$
This last relation is intuitively understood: 
for a large instability of the periodic orbit $g_{p}>>1$,
the resonances' lifetime $\tau_{a}=\hbar/\Gamma_{a}<<1$. 

Let us return on our general problem, from 2d to 3d. 
Let us comment that resonances 
will not always dominate the time evolution
of a wavepacket.
In fact, in a system like our 
one, one could expect
so many resonances
that after the first decays 
the system will proceed
to an average distribution over these
resonances' peaks.  
Considering a wavepacket $\psi_{t}(\vect{r})$
over many resonances in a region $W$ in the $\nu$-dimensional space,
the quantum survival probability is 
\be \label{intovr}
P(t)=\int_{W}|\psi_{t}({\bf r})|^{2}d\vect{r}
\ee
that can be also rewritten in terms of the initial density 
operator $\rho_{0}=|\psi_{0}\rangle \langle \psi_{0}|$
as 
\be \label{intial}
P(t)=tr \mathcal{I}_{D}({\bf r})e^{-\frac{iHt}{\hbar}}\rho_{0}e^{+\frac{iHt}{\hbar}}
\ee
where $\mathcal{I}_{D}$ is a distribution equal to $1$ for ${\bf r}$
into $D$ while is zero out of the region $D$. 
As done for the time-delay, one can 
express the survival probability in a semiclassical form
\be \label{survsemicl}
P(t)\simeq \int \frac{d\Gamma_{ph}}{(2\pi \hbar)^{f}}\mathcal{I}_{D}e^{{\bf L}_{cl}t}\tilde{\rho}_{0}+O(\hbar^{-\nu+1})+\frac{1}{\pi \hbar}\int dE\sum_{p}\sum_{a}\frac{\cos\left(a\frac{S_{p}}{\hbar}-a\frac{\pi}{2}{\bf m}_{p} \right)}{\sqrt{|det({\bf m}_{p}^{a}-{\bf 1})|}}\int_{p}\mathcal{I}_{D}e^{{\bf L}_{cl}t}\tilde{\rho}_{0}dt +O(\hbar^{0})
\ee
 where ${\bf L}_{cl}$ is the classical Liouvillian operator,
 defined in terms of classical Poisson brackets as 
 ${\bf L}_{cl}=\{H_{cl},...\}_{Poisson}$; $\tilde{\rho}_{0}$
 is the Wigner transform 
 of the initial density state
 
 The Sturm-Liouville problem associated to ${\bf L}_{cl}$
 defines the Pollicott-Ruelle resonances 
 \be \label{LclSL}
 {\bf L}_{cl}\phi_{n}=\{H_{cl},\phi_{n} \}_{Poisson}=\lambda_{n}\phi_{n}
 \ee
The eigenstates $\phi_{n}$ are Gelfald-Schwartz distributions.
They are the ones with unstable manifolds in the invariant set. 
On the other hand, the adjoint problem 
 \be \label{LclSL}
 {\bf L}_{cl}^{\dagger}\tilde{\phi}_{n}=\tilde{\lambda}_{n}\tilde{\phi}_{n}
 \ee
has eigenstates associated to stable manifolds. 
The eigenvalues $\lambda_{n}$  
are in general complex.
They have a real part $Re(\lambda_{n})\leq 0$
because of they are associated to an ensamble bounded periodic orbits.
On the other hand $Im(\lambda_{n})$ describe 
the decays of the statistical ensambles.
One can expand 
the survival probability over the Pollicot-Ruelle resonances as 
\be \label{Pexp}
P(t)\simeq \int \sum_{n}\langle \mathcal{I}_{D}|\phi_{n}(E)\rangle \langle \tilde{\phi}_{n}(E)|e^{\lambda_{n}(E)t}| \phi_{n}(E)\rangle \langle \tilde{\phi}_{n}(E)|\tilde{\rho}_{0}\rangle 
\ee 
From this expansion, one can consider the 0-th leading order: 
it will be just proportional to an exponential $e^{\lambda_{0}(E)t}$.
The long-time decay of the system is
expected to be related to the classical escape rate $\gamma(E)$.
So that we conclude that the survival probability 
goes as $P(t)\sim e^{-\gamma(E)t}$, {\it i.e} $s_{0}=-\gamma(E)$. 

As a consequence, the cross sections from A to B $\sigma_{AB}=|S_{AB}|^{2}$
are dramatically controlled
by the Pollicott-Ruelle resonances. 
let us consider cross sections' autocorrelations
\be \label{autocc}
C_{E}(\bar{E})=\langle \sigma_{BA}(E-\frac{\bar{E}}{2}) \sigma_{AB}(E+\frac{\bar{E}}{2})\rangle-|\langle \sigma_{BA}(E)\rangle|^{2}
\ee 
with $E$ labelling the energy shell considered. 
Let us perform the Fourier transform 
\be \label{Ftcc}
\tilde{C}_{E}(t)=\int_{-\infty}^{+\infty}C_{E}(\bar{E})e^{-\frac{i}{\hbar}\bar{E}t}d\bar{E}
\ee
As done for the survival probability, we expand (\ref{Ftcc}) 
all over the Pollicott-Ruelle spectrum 
so that we obtain 
\be \label{CE}
\tilde{C}_{E}(t)\simeq \sum_{n} \tilde{C}_{n}exp(-Re\lambda_{n}(E)t)\cos Im \lambda_{n}(E)t
\ee
where $\tilde{C}_{n}$ are coefficients of this expansion. 
In particular the leading order of (\ref{CE}) is related to (\ref{Pexp})
for $Im \lambda_{0}=0$:
\be \label{related}
\tilde{C}_{E}(t)\simeq exp(-\gamma(E)t)
\ee
corresponding to the main Lorentzian peak 
\be \label{mainllrr}
C_{E}(\bar{E})\sim \frac{1}{\bar{E}^{2}+(\hbar \gamma(E))^{2}}
\ee
while (\ref{CE})
corresponds to 
a spectral correlation 
\be \label{summm}
C_{E}(\bar{E})\simeq \sum_{n}\left\{ \frac{C_{n}}{(\bar{E}-\hbar Im\lambda_{n})^{2}+(\hbar Re \lambda_{n})^{2}}+
 \frac{C_{n}}{(-\bar{E}-\hbar Im\lambda_{n})^{2}+(\hbar Re \lambda_{n})^{2}}\right\}
\ee

We conclude resuming that a semiclassical quantum chaotic scattering approach 
leads to following conclusions about the box of geometries problem:
i) the existence of chaotic regions of trapped trajectories has to be a consequence 
of our scattering problem;
ii) the qualitative behavior of survival probability
and 
 correlation function is qualitatively understood as a decreasing
 function in time with an exponent determined by classical chaos scattering considerations.

\subsubsection{Quantum field theories}

In this section, we will formally discuss the problem 
of scattering from a QFT point of view.
This is based on the path integral approach on N geometries.
In the path integral integration, one will start to "explore" 
field configurations with energies comparable to the 
inverse asperities' size. In the fourier transform space, 
field configurations will be prevalently 
scattered by asperities if the 
a system has energy comparable to the inverse asperities' size
We are against a chaotic quantum field theory problem.
In a chaotic quantum field theory, 
there are not trapped trajectories in space-time 
but there trapped configurations in the 
infinite dimensional space of fields!
In analogy to semiclassical chaotic non-relativistic quantum mechanics, 
one can consider a semiclassical approximation in a regime in which 
the fields' action is much higher than $\hbar$:
$I>>\hbar$. 
In this approximation, we have a formal understanding of the 
chaotic quantum field theory problem. 
The corresponding WKB propagator
for a quantum field has a formal expression
\be \label{WKB}
\langle \phi_{0},t_{0}| \phi_{1},t_{1}\rangle\simeq \sum_{n}\mathcal{A}_{n}(\phi_{0},t_{0}|\phi_{1},t_{1})e^{\frac{i}{\hbar}I_{n}}
\ee
where we are summing on all over the classical orbits 
in the fields' configurations' space, 
while amplitudes $\mathcal{A}_{n}$
are 
\be \label{amplitudesss}
\mathcal{A}_{n}(\phi_{0},t_{0}|\phi_{1},t_{1})=\frac{1}{(2\pi i \hbar)^{\nu/2}}\sqrt{|det[\partial_{\phi_{0}}\partial_{\phi_{0}}I_{n}[\vect{r},\vect{r}_{0},t]]|}e^{-\frac{i\pi h_{n}}{2}}
\ee
where $h_{n}$ counts the number of conjugate points along the n-th orbits. 

We will expect that all rigorous results obtained in literature 
of classical chaotic scatterings, about the existence of invariant set 
with their topological robust proprieties discussed in part above, 
will be not rigorously extended for an infinite dimensional space of fields.
A complete theory regarding these aspects in QFT is not known to me.
Neverthless let us intuitively think something similar
 happen in space of fields, even if more complicated.
 The presence of chaotic zones of trapped periodic fields' configurations 
in a subregion of the configurations' space, corresponding 
to the one confined into our system, is expected for our problem.
Also for fields, chaotic unstable trajectories in the fields' space
are expected, as well as a large number of fields' resonances
in QFT S-matrices, generalizing Pollicot-Ruelle ones. 
The survival probability for a field are expected to 
exponentially decrease as in semiclassical quantum mechanical case. 

On the other hand, a general space of different fields,
the presence of interaction terms in the lagrangian leads
to tree-level transitions' processes
that has to be considered as leading orders in the
semiclassical saddle point perturbative expansion. 
As a consequence, chaotic fields' trapped trajectories
have to be thought as a multifields' ones. 
The result can be imagined as a 
chaotic cascade of processes among fields,
in which a part of different fields are trapped 
in the system, interacting
and scatterings and decaying each others.
For example, let us imagine one pure electromagnetic wave
entering inside the box of geometries. 
This starts to be diffracted into 
different direction,
so that initial coherent photons 
will start to re-meet each other in 
a different state. 
Of course, if their energy is enough, 
they can produce couples of $e^{+}e^{-}, q\bar{q}$ and so on. 
Then, these fields will interact each other through 
electromagnetic, strong and weak interactions.
The final system will be full of new fields, 
and it will have highly chaotic trapped zone.

An alternative formal way it the following. 
Suppose interdistances much higher than geometries' dimensions.
This case is a simplified one with respect to the realistic problem. 
In this case, we can define a transition amplitude for each geometry.
Let us suppose to be interested to calculate the transition amplitude for 
a field configuration $\phi_{0}$ to a field configuration $\phi_{N}$.
$\phi_{0}$ is the initial field configuration defined on a $t_{0}$, 
before entering in the system, while $\phi_{N}$ is a field configuration 
of a time $t_{N}$, corresponding to a an out-going state from the
system.
For simplicity, we can formalize the simplified problem as a 4D-box,
with $n \times m \times p$ 
 singularities in 3D, $n$ in the x-axis, $m$ in y-axis, $p$ in z-axis
(not necessary disposed as a regular lattice).  
Let us call $\mathcal{N}_{1},\mathcal{N}_{2}$ the sides sited in the xy-planes, 
$\mathcal{M}_{1,2}$ in xz-planes, $\mathcal{P}_{1,2}$ in zy-planes, 
delimiting the 3D-space-box. 
Let us consider an incident field $\phi_{0}$
on the 2D plane $\mathcal{N}_{1}$,
with $n\times m$ singularities. 
Then the $n\times m$  singularities will scatter the 
incident field in $n\times m$-waves.
From each diffractions, the out-waves will scatter on a successive 
singularity, penetrating in the box, 
or to the other nodes in the same plane $\mathcal{N}_{1}$,
and so on. Our problem is to evaluate the S-matrix from the in-state 0 to the out-the box state. 
One will expect that a fraction of initial probability density
will escape from the 3D box by the sides $\mathcal{N}_{1,2}\mathcal{M}_{1,2},\mathcal{P}_{1,2}$, 
 another fraction will be trapped "forever" (for a time-life equal to the one of the system) inside the box. 
As a consequence, one has to consider all possible diffraction stories or diffraction paths. 
Clearly, one has also to consider paths in which the initial wave
goes back and forth in the system before going-out.

One example of propagation Path $0-111-222-333-...-nmp-N$ 
\be \label{examplepath}
\langle\phi_{0},t_{0}|\phi_{111,in},t_{111,in}\rangle \langle\phi_{111,in},t_{111,in}|\phi_{111,out},t_{111,out}\rangle   
\langle\phi_{111,out},t_{111,out}|\phi_{222,in},t_{222,in}\rangle
\ee
$$\times \langle\phi_{222,in},t_{222,in}|\phi_{222,out},t_{222,out}\rangle...\langle\phi_{(n-1,m-1,p-1)},t_{(n-1),(m-1),(p-1)}|\phi_{nmp},t_{nmp}\rangle\langle\phi_{n,m,p},t_{n,m,p}|\phi_{N},t_{N}\rangle$$
where $|\phi_{ijk,in},t_{ijk,in}\rangle$ and $|\phi_{ijk,out},t_{ijk,out}\rangle$
are states before and after entering in the horizonless geometry $ijk$. 
In order to evaluate $\langle \phi_{0},t_{0}|\phi_{nmp},t_{nmp}\rangle$  
 one has to consider all the possible propagation paths 
from the initial position to the $nmp$-th singularity
\footnote{Propagators inside singular curved geometries like 
horizonless conic singularities mathematically exist and they were 
discussed in papers \cite{conic}. Other discussions about scalar fields on Kasner space-time 
can be found in \cite{Battista:2014fea}.
Contrary to the definition of propagators 
in quantum space-time foams, in our case the N geometries are well defined and continuos.
As a consequence one can define the propagation of fields inside them. Problematic regions are discontinuous ones among the geometries. In principle, one has to define also propagators 
for these linking regions, with an opportune smoothing procedure 
applied on these regions.}.
We define these amplitudes as
\be \label{MinkoP}
\langle \phi_{ijk},t_{ijk}|\phi_{i'j'k',in},t_{i'j'k',in} \rangle= \int_{\mathcal{M}_{0}}\mathcal{D}\phi e^{iI[\phi]}
\ee
while 
\be \label{MinkoP}
\langle \phi_{ijk,in},t_{ijk,in}|\phi_{ijk,out},t_{ijk,out} \rangle= \int_{\mathcal{M}_{ijk}}\mathcal{D}\phi e^{iI[\phi]}
\ee
where $\mathcal{M}_{0}$ is the Minkowski space-time, while $\mathcal{M}_{ijk}$ is the ijk-cone space-time. 
Again one can easily get that for a large system of naked singularities, 
it will exist a class of propagators' paths,
reaching the out state $|\phi_{N}, t_{N} \rangle$ 
only for a time $t_{N}\rightarrow \infty$.
A simple example can be the propagator paths
\be \label{propath}
|\langle \phi_{ijk},t_{ijk}|\phi_{i'j'k'},t_{i'j'k'} \rangle|^{2} |\langle \phi_{ijk},t^{(1)}_{ijk}|\phi_{i'j'k'},t_{i'j'k'}^{(1)} \rangle|^{2} ....|\langle \phi_{ijk},t^{(\infty)}_{ijk}|\phi_{i'j'k'},t_{i'j'k'}^{(\infty)} \rangle|^{2}
\ee
where $t_{ijk}^{\infty}>....>t_{ijk}^{(1)}>t_{ijk}$ and $t_{i'j'k'}^{\infty}>....>t_{i'j'k'}^{(1)}>t_{i'j'k'}$.
This amplitude is non-vanishing in such a system as an infinite sample of other ones. 
We can formally group these propagators in 
a $\langle BOX|BOX \rangle$ propagator, evaluating the probability 
that a field will remain in the box of singularities after a time 
larger than the system life-time. 
On the other hand, let call $\langle BOX|OUT \rangle$
and $\langle OUT|OUT \rangle$
the other processes. 

Considering interactions, one will also use S-matrices. 
We can write a generic S-matrix for one diffraction path 
as \be \label{Sgeneric}
\langle in|S^{Kth}|out \rangle=S_{0-1jk}S_{ijk}S_{i'j'k'}.....S_{(i^{n-1}j^{m-1}k^{p-1})-(i^{n}j^{m}k^{p})}
\ee
A class of paths like 
with conditions 
\be \label{cond1}
i\leq i' \leq i+1
\ee
\be \label{cond2}
j\leq j' \leq j+1
\ee
\be \label{cond1}
k\leq k' \leq k+1
\ee
$$...$$
\be \label{cond1}
i^{n-1}\leq i^{n} \leq i^{n-1}+1
\ee
\be \label{cond2}
j^{m-1}\leq j^{m} \leq j^{m-1}+1
\ee
\be \label{cond1}
k^{p-1}\leq k^{p} \leq k^{p-1}+1
\ee
We call these class of paths "minimal paths". 
In fact, in these paths there are not 
back-transitions. 
The total number of "minimal paths" is 
is $n\times m \times p \times (n-1)$.
On the other hand, the number of paths 
with back and forth scatterings will diverge. 

As a consequence, the total S-matrix
is the sum over all possible infinite diffraction paths
\be \label{TOT}
\langle in|S^{OUT}_{n}|out \rangle=\sum_{paths}\langle in |S^{K-th}_{n}|out \rangle
\ee
accounting for all the paths leading from the in-state to 
the out-of-box state.

For a completeness of our discussion, 
let us reformulate the non-relativistic quantum problem 
in a non-relativistic path integral formulation.
We will use here the bracket-notation,
in which the propagator from $(x_{0},t_{0})$ to $(x_{1},t_{1})$ is
$$K(x_{0},t_{0};x,t_{1})=\langle x_{0},t_{0}| x_{1}, t_{1} \rangle$$
 
This will be equivalent to wave functions' formulation considered 
in section 3.2. 
In this case, a problem of $\langle OUT|OUT \rangle$ 
is reformulated not with propagators in the fields' space 
but in the same space-time points.
$\langle OUT|OUT \rangle$ will account for all possible 
paths leading to an in-coming state $|x_{0},t_{0}\rangle$
 to another state out of the box. 
 Again, such a problem is chaotized by the fact that 
 one has to consider the interference of all possible paths 
 passing for all possible horizonless geometries. 
 A simple example of a path inside the OUT-OUT class of paths 
 is $0-111-222-333-...-nmp-N$ 
\be \label{examplepath}
\langle x_{0},t_{0}|x_{111,in},t_{111,in}\rangle \langle x_{111,in},t_{111,in}|x_{111,out},t_{111,out}\rangle   
\langle x_{111,out},t_{111,out}|x_{222,in},t_{222,in}\rangle
\ee
$$\times \langle x_{222,in},t_{222,in}|x_{222,out},t_{222,out}\rangle...\langle x_{(n-1,m-1,p-1)},t_{(n-1),(m-1),(p-1)}|x_{nmp},t_{nmp}\rangle$$
where $|x_{ijk,in},t_{ijk,in}\rangle$ and $|x_{ijk,out},t_{ijk,out}\rangle$
are states before and after entering in the geometry $ijk$. 

One can find trapped propagators 
like 
\be \label{propath}
|\langle x_{ijk},t_{ijk}|x_{i'j'k'},t_{i'j'k'} \rangle|^{2} |\langle x_{ijk},t^{(1)}_{ijk}|x_{i'j'k'},t_{i'j'k'}^{(1)} \rangle|^{2} ....|\langle x_{ijk},t^{(\infty)}_{ijk}|x_{i'j'k'},t_{i'j'k'}^{(\infty)} \rangle|^{2}
\ee
where $t_{ijk}^{\infty}>....>t_{ijk}^{(1)}>t_{ijk}$ and $t_{i'j'k'}^{\infty}>....>t_{i'j'k'}^{(1)}>t_{i'j'k'}$.
A class of paths from OUT to BOX state will be attracted in these 
trapped paths. 


\section{Conclusions and outlooks}

In this paper, we have shown how 
a semiclassical black hole can be obtained as a system of naked singularities 
(even if probably smoothed at the Planck lenght). 
In particular, we have mathematically defined a new object called
frizzyball. 
A frizzyball emits a quasi-thermal radiation.
Deviations from thermality are related to 
the particular disposition of geometries, so that 
they carry small informations about the space-time structure.
In such a system, we have shown how
infalling wave functions will be inevitably chaotized.
These chaotic effects will manifest themself 
before ingoing into the full transplanckian regime. 
These statements are sustained from 
the analysis  
of
the non-relativistic scattering problem of a particle
ingoing to a system of N horizonless geometries.

This system is a Sinai Biliard of the space-time topology.
We have argumented how trapped chaotized 
zones will be formed among the space-time asperities. 
However, the frizzy topology is sustained by the 
frizzy-ball mass. But the mass is gradually lost in quasi Bekenstein-Hawking 
evaporation. As argumented, at a certain critical mass, 
a phase transition of the space-time topology 
into a trivial Minkowski vacuum state is expected. 
During this process, asperities are gradually washed-out
and trapped information are gradually leaked in the external environment. 
The final evaporation will release all hidden information. We have called this phenomena 
 "final information burst". As a consequence, 
the S-matrix $\langle collapse|S|total\, evaporation \rangle$
is unitary and well define, without any paradoxes. 
From the point of view of an external 
observer, for $M>M_{\odot}$ such a process will be longer than the age of the Universe
($t_{BH}>10^{74}\, \rm s$),
while from the reference frame of an internal trapped particle, 
the time will be very short $t_{BH}<<1\, \rm yr$ (depending by the frizzy-ball mass).  
A part of the information will be trapped by semiclassical black hole just because 
of the causal structure of the interior, so our proposal seems un-useful from this point of view.
However, there is an important difference, that can be understood as follows.
Let us consider an entangled Hawking's pair, one outgoing and the other one ingoing. 
In a semiclassical black hole, they will remain entangled and the smoothed causal structure
cannot disentangle them. On the other hand, if as suggested
the infalling one will start to be scattered in the Sinai biliard, 
then it will start to be very efficiently converted in a shower of particles, as shown above.
As a consequence, the initial entanglement is chaotically lost 
because of quantum decoherence: the out-going pair is 
entangled with a large number of fields inside the system.  
The efficiency of this process will exponentially increase with the number of 
infalling particles. This is a new quantum decoherence effect 
induced by the space-time topology. 
Let us note that the causal structure of a frizzyball
has a non-trivial topology.
As a consequence, the associated Penrose' diagram is 
a complicated superposition of Penrose' diagrams
of the different metrics. 

At this point, I am tempted to suggest that 
black holes and frizzyballs 
could be observationally differentiated
through their gravitational lensing proprieties. 
In  \cite{Virbhadra:1998dy,Virbhadra:1999nm}, the difference between gravitational lensing signatures of 
black holes and naked singularities were discussed in details.
In a broad sense, frizzyballs are new solutions interpolating 
among black holes' and naked singularities' ones.
In fact, the asperity parameter $\sigma_{\beta}^{2}$ 
defined in section 2 is connected to geometric standard deviations of the frizzyballs'
surfaces $\sigma_{r}^{2}$ ($r$ bh radius) with respect to a (semi)classical black hole.
In our model, $\sigma_{r}$ is a free parameter in a large range 
$l_{Pl}<\sigma_{r}<r_{S}$ ($l_{Pl}$ is the Planck lenght, $r_{S}$ the Schwarzschild radius).  
$\sigma_{r}\leq l_{pl}$ corresponds to a black hole while $\sigma_{r} \geq r_{S}$ to 
separated naked singularities. 
However, in the framework of our semiclassical effective model, 
we cannot establish the largeness of such a parameter.  
If such a parameter is determined by
an UV completion of our model and/or if it depends 
on initial conditions of star collapses is still unclear. 
In fact, these problems could be connected to other 
deep issues. 
For instance, the validity of 
 (weak or strong) cosmic censorship 
conjectures and hoop conjectures still remains unclear.
In fact, see references \cite{cosmic} for an overview of discussions on the cosmic censorship
conjectures and their possible violations;
 see references \cite{hoop} about the hoop conjecture. 
 If frizzyballs exist, these conjectures will have to be considered 
 true only as {\it approximated/accidental} ones. 
Anyway, the dynamical reason of such conjectures remains
unknown. 
A possibility could be that  frizzyball could be formed through a {\it dynamical 
censorship mechanism}. We define {\it dynamical censorship mechanism}
as a generic evolution from a configuration of separated naked singularities to 
a frizzyball.  Separated naked singularities could be 
destabilized by external electromagnetic and/or gravitational and/or matter fields'
perturbations. These kinds of instabilities at classical level,
for Super-extremal Kerr naked singularities $J^{2}>M>0$
($J$ bh spin parameter, $M$ bh mass)
and Super-extremal Reissner-N$\ddot{o}$rdstrom
bh $|Q|>M>0$ ($Q$ bh charge),
were discussed in 
\cite{Dotti}.  
Because of that, formation of frizzyballs could be energetically convenient,
as (meta)stable configurations against perturbations.  
Evidences from numerical simulations, that naked singularities 
can be formed in collapses  \cite{numerical}, seem to sustain our hypothesis. 
If an unstable naked singularity was formed, it would decay to  
a system of naked singularities disposed as a frizzyball.   
As a consequence, a frizzyball with $\sigma_{r}>>l_{Pl}$ could be 
detectable through its gravitational lensing characteristics.
These aspects deserve future studies beyond the purpose
of this paper.

To conclude, the black hole interior could be frizzy rather 
than smoothed as thought in semiclassical approach,
even at smaller energy scales compared to the Planck scale. 
This 
will induce a high chaotization of infalling information.
We are not proposing a final solution about BH problems, 
but a possible different point of view on these issues. 

I think that this hypothesis deserves future investigations 
by different communities of physicists, from chaos theory
and (classical and quantum) gravity theory \footnote{
For other discussions on these aspects see \cite{Addazi:2015hpa}
.  }.

\begin{acknowledgments} I would like to thank 
Gia Dvali, Gerard t'Hooft, Carlo Rovelli, Salvatore Capozziello, Giampiero Esposito,
Kumar Shwetketu Virbhadra, Leopoldo A. Pando Zayas,
Dejan Stojkovic, Sam Braunstein and Silvia Vada 
 for interesting conversations on these subjects. 
 I also would like to thank organizers of 14th Marcell Grossmann 
 meeting in Rome and Karl Schwarzschild meeting 2015 in Frankfurt
 for hospitality during the preparation of this paper. 
My work was supported in part by the MIUR research
grant "Theoretical Astroparticle Physics" PRIN 2012CPPYP7.

\end{acknowledgments}


\vspace{0.5cm}
\begin{thebibliography}{99}
  
\bibitem{Bekenstein}
J. D. Bekenstein, Phys. Rev. {\bf D7} (1973) 2333.  
  
\bibitem{Hawking}
S. W. Hawking, 
Phys. Rev. D {\bf 13}, 2 (1976).

\bibitem{Maldacena:1997re}
  J.~M.~Maldacena,
  Int.\ J.\ Theor.\ Phys.\  {\bf 38} (1999) 1113
   [Adv.\ Theor.\ Math.\ Phys.\  {\bf 2} (1998) 231]
  [hep-th/9711200].

\bibitem{Witten:1998qj}
  E.~Witten,
  Adv.\ Theor.\ Math.\ Phys.\  {\bf 2} (1998) 253
  [hep-th/9802150].
  
\bibitem{Gubser:1998bc}
  S.~S.~Gubser, I.~R.~Klebanov and A.~M.~Polyakov,
  Phys.\ Lett.\ B {\bf 428} (1998) 105
  [hep-th/9802109].
  

\bibitem{Susskind:1993if}
  L.~Susskind, L.~Thorlacius and J.~Uglum,
  Phys.\ Rev.\ D {\bf 48} (1993) 3743
  [hep-th/9306069].
  
\bibitem{'tHooft:1984re}
  G.~'t Hooft,
  Nucl.\ Phys.\ B {\bf 256} (1985) 727.
  
\bibitem{Braunstein:2009my}
  S.~L.~Braunstein, S.~Pirandola and K.~?yczkowski,
  Phys.\ Rev.\ Lett.\  {\bf 110} (2013) 10,  101301
  [arXiv:0907.1190 [quant-ph]].
  
\bibitem{Almheiri:2012rt}
  A.~Almheiri, D.~Marolf, J.~Polchinski and J.~Sully,
  JHEP {\bf 1302} (2013) 062
  [arXiv:1207.3123 [hep-th]].
  
  \bibitem{Fuzz1}
  O. Lunin and S. D. Mathur, Nucl. Phys. B {\bf 623}, 342 (2002) [arXiv:hep-th/0109154];
  O. Lunin and S. D. Mathur, Phys. Rev. Lett. {\bf 88}, 211303 (2002) [arXiv:hep-th/0202072];
O. Lunin, J. Maldacena and L. Maoz, [arXiv:hep-th/0212210];
O. Lunin and S. D. Mathur, Nucl. Phys. B {\bf 615}, 285 (2001) [arXiv:hep-th/0107113];
S. D. Mathur, A. Saxena and Y. K. Srivastava, Nucl. Phys. B {\bf 680}, 415 (2004)
[arXiv:hep-th/0311092];
O. Lunin, JHEP {\bf 0404}, 054 (2004) [arXiv:hep-th/0404006];
S. Giusto, S. D. Mathur and A. Saxena, Nucl. Phys. B {\bf 701}, 357 (2004)
[arXiv:hep-th/0405017];
S. Giusto, S. D. Mathur and A. Saxena, Nucl. Phys. B {\bf 710}, 425 (2005)
[arXiv:hep-th/0406103];
I. Bena and N. P. Warner, Adv. Theor. Math. Phys. {\bf 9}, 667 (2005) [arXiv:hep-th/0408106];
V. Jejjala, O. Madden, S. F. Ross and G. Titchener, Phys. Rev. D {\bf 71}, 124030 (2005)
[arXiv:hep-th/0504181];
I. Bena and N. P. Warner, Phys. Rev. D {\bf 74}, 066001 (2006) [arXiv:hep-th/0505166];
P. Berglund, E. G. Gimon and T. S. Levi, JHEP {\bf 0606}, 007 (2006) [arXiv:hep-th/0505167];
M. Taylor, JHEP {\bf 0603}, 009 (2006) [arXiv:hep-th/0507223];
A. Saxena, G. Potvin, S. Giusto and A. W. Peet, JHEP {\bf 0604}, 010 (2006)
[arXiv:hep-th/0509214];
I. Bena, C. W. Wang and N. P. Warner, Phys. Rev. D {\bf 75}, 124026 (2007)
[arXiv:hep-th/0604110];
V. Balasubramanian, E. G. Gimon and T. S. Levi, JHEP {\bf 0801}, 056 (2008)
[arXiv:hep-th/0606118];
I. Bena, C. W. Wang and N. P. Warner, JHEP {\bf 0611}, 042 (2006) [arXiv:hep-th/0608217];
K. Skenderis and M. Taylor, Phys. Rev. Lett. {\bf 98}, 071601 (2007) [arXiv:hep-th/0609154];
J. Ford, S. Giusto and A. Saxena, Nucl. Phys. B {\bf 790}, 258 (2008) [arXiv:hep-th/0612227];
I. Bena and N. P. Warner, Lect. Notes Phys. {\bf 755}, 1 (2008) [arXiv:hep-th/0701216];
I. Bena, N. Bobev and N. P. Warner, JHEP {\bf 0708}, 004 (2007) [arXiv:0705.3641 [hep-th]];
E. G. Gimon and T. S. Levi, JHEP {\bf 0804}, 098 (2008) [arXiv:0706.3394 [hep-th]];
I. Bena, C. W. Wang and N. P. Warner, JHEP {\bf 0807}, 019 (2008) [arXiv:0706.3786 [hepth]];
V. Cardoso, O. J. C. Dias and R. C. Myers, Phys. Rev. D {\bf 76}, 105015 (2007)
[arXiv:0707.3406 [hep-th]];
S. Giusto, S. F. Ross and A. Saxena, JHEP {\bf 0712}, 065 (2007) [arXiv:0708.3845 [hep-th]];
B. D. Chowdhury and S. D. Mathur, Class. Quant. Grav. {\bf 25}, 135005 (2008)
[arXiv:0711.4817 [hep-th]];
K. Skenderis and M. Taylor, Phys. Rept. {\bf 467}, 117 (2008) [arXiv:0804.0552 [hep-th]];
I. Bena, N. Bobev, C. Ruef and N. P. Warner, arXiv:0804.4487 [hep-th];
B. D. Chowdhury and S. D. Mathur, arXiv:0806.2309 [hep-th];
J. de Boer, S. El-Showk, I. Messamah and D. Van den Bleeken, JHEP {\bf 0905}, 002 (2009)
[arXiv:0807.4556 [hep-th]];
B. D. Chowdhury and S. D. Mathur, Class. Quant. Grav. {\bf 26}, 035006 (2009)
[arXiv:0810.2951 [hep-th]];
V. Balasubramanian, J. de Boer, S. El-Showk and I. Messamah, Class. Quant. Grav. {\bf 25},
214004 (2008) [arXiv:0811.0263 [hep-th]];
I. Bena, N. Bobev, C. Ruef and N. P. Warner, JHEP {\bf 0907}, 106 (2009) [arXiv:0812.2942
[hep-th]];
S. G. Avery and B. D. Chowdhury, arXiv:0907.1663 [hep-th];
J. H. Al-Alawi and S. F. Ross, JHEP {\bf 0910}, 082 (2009) [arXiv:0908.0417 [hep-th]];
I. Bena, S. Giusto, C. Ruef and N. P. Warner, JHEP {\bf 0911}, 089 (2009) [arXiv:0909.2559
[hep-th]];
N. Bobev and C. Ruef, JHEP {\bf 1001}, 124 (2010) [arXiv:0912.0010 [hep-th]];
S. G. Avery, B. D. Chowdhury and S. D. Mathur, arXiv:0906.2015 [hep-th];
S. Giusto, J. F. Morales and R. Russo, JHEP {\bf 1003}, 130 (2010) [arXiv:0912.2270 [hep-th]].

  
\bibitem{Dvali:2011aa}
  G.~Dvali and C.~Gomez,
  Fortsch.\ Phys.\  {\bf 61} (2013) 742
  [arXiv:1112.3359 [hep-th]].

\bibitem{Dvali:2012gb}
  G.~Dvali and C.~Gomez,
  Phys.\ Lett.\ B {\bf 716} (2012) 240
  [arXiv:1203.3372 [hep-th]].

\bibitem{Dvali:2012rt}
  G.~Dvali and C.~Gomez,
  Phys.\ Lett.\ B {\bf 719} (2013) 419
  [arXiv:1203.6575 [hep-th]].

\bibitem{Dvali:2012en}
  G.~Dvali and C.~Gomez,
  Eur.\ Phys.\ J.\ C {\bf 74} (2014) 2752
  [arXiv:1207.4059 [hep-th]].
  
\bibitem{Dvali:2014ila}
  G.~Dvali, C.~Gomez, R.~S.~Isermann, D.~L$\ddot{u}$st and S.~Stieberger,
  Nucl.\ Phys.\ B {\bf 893} (2015) 187
  [arXiv:1409.7405 [hep-th]].
  
\bibitem{Dvali:2015ywa}
  G.~Dvali, A.~Franca, C.~Gomez and N.~Wintergerst,
  arXiv:1507.02948 [hep-th].
  
\bibitem{Foit:2015wqa}
  V.~F.~Foit and N.~Wintergerst,
  arXiv:1504.04384 [hep-th].
  
\bibitem{Casadio:2015bna}
  R.~Casadio, A.~Giugno and A.~Orlandi,
  Phys.\ Rev.\ D {\bf 91} (2015) 12,  124069
  [arXiv:1504.05356 [gr-qc]].
  
\bibitem{Casadio:2015sda}
  R.~Casadio, O.~Micu and D.~Stojkovic,
  Phys.\ Lett.\ B {\bf 747} (2015) 68
  [arXiv:1503.02858 [gr-qc]].
  

\bibitem{Hooft:2014daa}
  G.~T.~Hooft,
  arXiv:1410.6675 [gr-qc].

\bibitem{Carr:2015nqa}
  B.~J.~Carr, J.~Mureika and P.~Nicolini,
  JHEP {\bf 1507} (2015) 052
  [arXiv:1504.07637 [gr-qc]].
  
\bibitem{Lake:2015pma}
  M.~J.~Lake and B.~Carr,
  arXiv:1505.06994 [gr-qc].
  
\bibitem{Barrau:2014yka}
  A.~Barrau, C.~Rovelli and F.~Vidotto,
  Phys.\ Rev.\ D {\bf 90} (2014) 12,  127503
  [arXiv:1409.4031 [gr-qc]].
    
\bibitem{Barrau:2015uca}
  A.~Barrau, B.~Bolliet, F.~Vidotto and C.~Weimer,
  arXiv:1507.05424 [gr-qc].
  
\bibitem{Rovelli:2014cta}
  C.~Rovelli and F.~Vidotto,
  Int.\ J.\ Mod.\ Phys.\ D {\bf 23} (2014) 12,  1442026
  [arXiv:1401.6562 [gr-qc]].
  
  
\bibitem{Saueressig:2015xua}
  F.~Saueressig, N.~Alkofer, G.~D'Odorico and F.~Vidotto,
  arXiv:1503.06472 [hep-th].
  
\bibitem{DeLorenzo:2014pta}
  T.~De Lorenzo, C.~Pacilio, C.~Rovelli and S.~Speziale,
  Gen.\ Rel.\ Grav.\  {\bf 47} (2015) 4,  41
  [arXiv:1412.6015 [gr-qc]].
  
\bibitem{Giddings:2014nla}
  S.~B.~Giddings,
  Phys.\ Lett.\ B {\bf 738} (2014) 92
  [arXiv:1401.5804 [hep-th]].
  
\bibitem{Giddings:2012gc}
  S.~B.~Giddings,
  Phys.\ Rev.\ D {\bf 88} (2013) 064023
  [arXiv:1211.7070 [hep-th]].
 
\bibitem{Germani:2015tda}
  C.~Germani and D.~Sarkar,
  arXiv:1502.03129 [hep-th].
  
\bibitem{Marolf:2015dia}
  D.~Marolf and J.~Polchinski,
  arXiv:1506.01337 [hep-th].
  
\bibitem{Chatwin-Davies:2015hna}
  A.~Chatwin-Davies, A.~S.~Jermyn and S.~M.~Carroll,
  arXiv:1507.03592 [hep-th].
  
\bibitem{Susskind:2015toa}
  L.~Susskind,
  arXiv:1507.02287 [hep-th].
  
\bibitem{Afshordi:2015foa}
  N.~Afshordi and Y.~K.~Yazdi,
  arXiv:1502.01023 [astro-ph.HE].
  
\bibitem{Grudka:2015wva}
  A.~Grudka, M.~J.~W.~Hall, M.~Horodecki, R.~Horodecki, J.~Oppenheim and J.~A.~Smolin,
  arXiv:1506.07133 [hep-th].
  
\bibitem{Firouzjaee:2015bqa}
  J.~T.~Firouzjaee and G.~F.~R.~Ellis,
  Phys.\ Rev.\ D {\bf 91} (2015) 10,  103002
  [arXiv:1503.05020 [gr-qc]].
  
      
\bibitem{Hawking:2014tga}
  S.~W.~Hawking,
  arXiv:1401.5761 [hep-th].

\bibitem{Kawai:2014afa}
  H.~Kawai and Y.~Yokokura,
  Int.\ J.\ Mod.\ Phys.\ A {\bf 30} (2015) 15,  1550091
  [arXiv:1409.5784 [hep-th]].
  
\bibitem{Kawai:2013mda}
  H.~Kawai, Y.~Matsuo and Y.~Yokokura,
  Int.\ J.\ Mod.\ Phys.\ A {\bf 28} (2013) 1350050
  [arXiv:1302.4733 [hep-th]].
  
 \bibitem{C0}
  S. H. Shenker and D. Stanford, Black holes and the butterfly effect, JHEP {\bf 1403} (2014) 067
[1306.0622];
  S. H. Shenker and D. Stanford, JHEP {\bf 12} (2014) 046 [1312.3296]; 
  S. H. Shenker and D. Stanford, JHEP {\bf 05} (2015) 132
[1412.6087];
J.~Maldacena, S.~H.~Shenker and D.~Stanford,
  arXiv:1503.01409 [hep-th].
  L.~A.~Pando Zayas and C.~A.~Terrero-Escalante,
  JHEP {\bf 1009} (2010) 094
  [arXiv:1007.0277 [hep-th]];
  A.~Farahi and L.~A.~Pando Zayas,
  Phys.\ Lett.\ B {\bf 734} (2014) 31
  [arXiv:1402.3592 [hep-th]];
   L.~A.~Pando Zayas,
  Int.\ J.\ Mod.\ Phys.\ D {\bf 23} (2014) 12,  1442013
  [arXiv:1405.3655 [hep-th]].
  
  
\bibitem{Polchinski:2015cea}
  J.~Polchinski,
  arXiv:1505.08108 [hep-th].
  
  
  \bibitem{C1}
 D.Marolf,  arXiv:1508.00939.
  
  \bibitem{QuantumChaos}
F. Haake,   {\it Quantum Signatures of Chaos} Edition: 2, Springer, 2001, ISBN 3-540-67723-2, ISBN 978-3-540-67723-9;
M.Berry, {\it Quantum Chaology}, pp 104-5 of {\it Quantum: a guide for the perplexed} by J-Al Khalili.

\bibitem{QCS}
T. T\'el, M. Gruinz, {\it Chaotic Dynamics}, Cambridge University Press (2006), Cambridge UK;
J.M. Seoane and M. A. F. Sanju\'an 2013 Rep. Prog. Phys. {\bf 76} 016001; 

\bibitem{QCS2}
V.I. Arnold, Russ.Math.Surv. 18:6:85-191;
J.K.Moser, Mem.Am.Math.Soc. {\bf 81}: 1-60.

\bibitem{QCS3}
B.Eckhardt, J.Phys.A: Math. Gen. {\bf 20}: 5971-5979.
P.Gaspard and S.A. Rice, J.Chem.Phys. {\bf 90}:2225-2241;
P.Gaspard, {\it Chaos, Scattering and Statistical Mechanics},
Cambridge University Press, Cambridge UK. 

\bibitem{QCS4}
K.T. Hansen, Nonlinearity {\bf 6}: 753-770;
M.M. Sano, J.Phys. A: Math. Gen. {\bf 27}: 4791-4803. 


\bibitem{Horodecki:2009zz}
  R.~Horodecki, P.~Horodecki, M.~Horodecki and K.~Horodecki,
  Rev.\ Mod.\ Phys.\  {\bf 81} (2009) 865
  [quant-ph/0702225].
  
          
\bibitem{Addazi:2014ila}
  A.~Addazi and M.~Bianchi,
  JHEP {\bf 1412} (2014) 089
  [arXiv:1407.2897 [hep-ph]].

\bibitem{Addazi:2014mga}
  A.~Addazi and S.~Capozziello,
  Int.\ J.\ Theor.\ Phys.\  {\bf 54} (2015) 6,  1818
  [arXiv:1407.4840 [gr-qc]].

\bibitem{Addazi:2015ata}
  A.~Addazi,
  JHEP {\bf 1504} (2015) 153
  [arXiv:1501.04660 [hep-ph]].

\bibitem{Addazi:2015rwa}
  A.~Addazi and M.~Bianchi,
  arXiv:1502.01531 [hep-ph].

\bibitem{Addazi:2015dxa}
  A.~Addazi and G.~Esposito,
  Int.\ J.\ Mod.\ Phys.\ A {\bf 30} (2015) 1550103
  [arXiv:1502.01471 [hep-th]].
  
\bibitem{Addazi:2015hka}
  A.~Addazi and M.~Bianchi,
  JHEP {\bf 1506} (2015) 012
  [arXiv:1502.08041 [hep-ph]].
  
\bibitem{Addazi:2015eca}
  A.~Addazi,
  arXiv:1504.06799 [hep-ph].

\bibitem{Addazi:2015fua}
  A.~Addazi,
  arXiv:1505.00625 [hep-ph].
  
\bibitem{Addazi:2015oba}
  A.~Addazi,
  arXiv:1505.02080 [hep-ph].

\bibitem{Addazi:2015ppa}
  A.~Addazi,
  arXiv:1505.07357 [hep-th].
  

\bibitem{Addazi:2015goa}
  A.~Addazi,
  arXiv:1506.06351 [hep-ph].
  
\bibitem{Addazi:2015yna}
  A.~Addazi, M.~Bianchi and G.~Ricciardi,
  arXiv:1510.00243 [hep-ph].
  
\bibitem{Addazi:2015ewa}
  A.~Addazi,
  arXiv:1510.02911 [hep-ph].
  
\bibitem{Addazi:2015pia}
  A.~Addazi,
  Nuovo Cim.\ C {\bf 038} (2015) 01,  21.
  
  
  \bibitem{Euclidean}
  G. W. Gibbons, S. W. Hawking, M. J. Perry, Nucl. Phys. B {\bf 138} 141 (1978);
   J. Hartle, K.Schleich, Phys. Rev. D {\bf 36} 2342 (1987).
   
   \bibitem{conic}
   D.V. Fursaev, S. N. Solodukhin. 
    arXiv preprint hep-th/9501127 (1995);
   D.V. Fursaev
    Nuclear Physics B {\bf 524}.1 (1998): 447-468;
    D.V. Fursaev, Dmitri V. 
    Physics Letters B {\bf 334}.1 (1994): 53-60;
    D.V. Fursaev, G. Miele. "Cones, spins and heat kernels." 
    Nuclear Physics B {\bf 484}.3 (1997): 697-723.
    

  
\bibitem{Virbhadra:1998dy}
  K.~S.~Virbhadra, D.~Narasimha and S.~M.~Chitre,
  Astron.\ Astrophys.\  {\bf 337} (1998) 1
  [astro-ph/9801174].
      
\bibitem{Virbhadra:1999nm}
  K.~S.~Virbhadra and G.~F.~R.~Ellis,
  Phys.\ Rev.\ D {\bf 62} (2000) 084003
  [astro-ph/9904193].
  
  
\bibitem{Virbhadra:2007kw}
  K.~S.~Virbhadra and C.~R.~Keeton,
  Phys.\ Rev.\ D {\bf 77} (2008) 124014
  [arXiv:0710.2333 [gr-qc]].
  
  
    

\bibitem{Battista:2014fea}
  E.~Battista, E.~Di Grezia and G.~Esposito,
  Int.\ J.\ Geom.\ Meth.\ Mod.\ Phys.\  {\bf 12} (2015) 1550060
  [arXiv:1410.3971 [gr-qc]].
  
  \bibitem{cosmic}
  R.Penrose,  "The Question of Cosmic Censorship", Chapter 5 in Black Holes and Relativistic Stars, Robert Wald (editor), (1994) (ISBN 0-226-87034-0);
R.Penrose, "Singularities and time-asymmetry", Chapter 12 in General Relativity: An Einstein Centenary Survey (Hawking and Israel, editors), (1979), see especially section 12.3.2, pp. 617-629 (ISBN 0-521-22285-0);
S.L. Shapiro, and S.A. Teukolsky,  
Physical Review Letters {\bf 66}, 994-997 (1991);
R.Wald, General Relativity, 299-308 (1984) (ISBN 0-226-87033-2);
J. Earman Bangs, Crunches, "Whimpers, and Shrieks: Singularities and Acausalities in Relativistic Spacetimes" (1995), see especially chapter 2 (ISBN 0-19-509591-X)
M.D. Roberts,
Gen.Rel.Grav.{\bf 21}(1989)907-939;
  K.~S.~Virbhadra,
  Phys.\ Rev.\ D {\bf 60} (1999) 104041
  [gr-qc/9809077].

  
  \bibitem{hoop}
  K S Thorne,  "Nonspherical Gravitational Collapse: A Short Review in
Magic without Magic ", ed. J Klauder (San Francisco: Freeman) (1972);
 K.S. Thorne,
  "Black Holes and Time Warps: Einstein's Outrageous Legacy", New York: W. W. Norton, 1994;
  J. M. M. Senovilla, 
  Europhys.Lett. {\bf 81}, 20004 (2008) [arXiv:0709.0695 [gr-qc]];
   G.~W.~Gibbons,
  arXiv:0903.1580 [gr-qc];
  T. Chiba, T. Nakamura, K. i. Nakao and M. Sasaki, Class. Quant. Grav. {\bf 11}, 431 (1994);
H. Yoshino, Y. Nambu and A. Tomimatsu, Phys. Rev. D {\bf 65}, 064034 (2002)
[arXiv:gr-qc/0109016];
T. Nakamura, S. L. Shapiro and S. A. Teukolsky, Phys. Rev. D {\bf 38}, 2972 (1988).
  
  \bibitem{numerical}
  D.~M.~Eardley and L.~Smarr,
  Phys.\ Rev.\ D {\bf 19} (1979) 2239;
   D.~M.~Eardley,
  NATO Sci.\ Ser.\ B {\bf 156} 229.
  P. S. Joshi and D. Malafarina, Phys Rev D{\bf  83,} 024009 (2011);
   P. S. Joshi and D. Malafarina, Gen. Rel. Grav. {\bf 45} (2), 305 (2013);
   S.~Satin, D.~Malafarina and P.~S.~Joshi,
  arXiv:1409.0505 [gr-qc];
    L.~Kong, D.~Malafarina and C.~Bambi,
  Eur.\ Phys.\ J.\ C {\bf 74} (2014) 2983
  [arXiv:1310.8376 [gr-qc]];
   P.~S.~Joshi, D.~Malafarina and R.~Narayan,
  Class.\ Quant.\ Grav.\  {\bf 31} (2014) 015002
  [arXiv:1304.7331 [gr-qc]];
  N.~Ortiz,
  AIP Conf.\ Proc.\  {\bf 1473} (2012) 49
  [arXiv:1204.4481 [gr-qc]];
  P.~S.~Joshi and D.~Malafarina,
  Int.\ J.\ Mod.\ Phys.\ D {\bf 20} (2011) 2641
  [arXiv:1201.3660 [gr-qc]];
   U.~Miyamoto, S.~Jhingan and T.~Harada,
  arXiv:1108.0248 [gr-qc].
  
  
    
  \bibitem{Dotti}
  R. J. Gleiser and G. Dotti, Class. Quant. Grav. {\bf 23}, 5063 (2006) [arXiv:gr-qc/0604021];
G. Dotti, R. Gleiser and J. Pullin, Phys. Lett. B {\bf 644}, 289 (2007)
[arXiv:gr-qc/0607052];
  G.~Dotti, R.~J.~Gleiser, I.~F.~Ranea-Sandoval and H.~Vucetich,
  Class.\ Quant.\ Grav.\  {\bf 25} (2008) 245012
  [arXiv:0805.4306 [gr-qc]];
  G.~Dotti and R.~J.~Gleiser,
  Class.\ Quant.\ Grav.\  {\bf 26} (2009) 215002
  [arXiv:0809.3615 [gr-qc]].

\bibitem{Addazi:2015hpa}
  A.~Addazi,
  arXiv:1510.05876 [gr-qc].

\end{thebibliography}
\end{document}